\pdfoutput=1

\documentclass[11pt]{article}

\usepackage{ACL2023}

\usepackage{times}
\usepackage{latexsym}

\usepackage[T1]{fontenc}

\usepackage{color}
\usepackage{array}
\usepackage{enumitem}
\usepackage{caption}
\usepackage{algorithm}
\usepackage{algorithmic}
\usepackage{amsmath}
\usepackage{amssymb}
\usepackage{graphicx}
\usepackage{multirow}
\usepackage{booktabs}
\usepackage{arydshln}
\usepackage{bm}
\usepackage{tikz}
\usepackage{pgfplots}
\usepackage{xspace}
\usepackage{paralist}
\usepackage{tabularx}
\usepackage{graphicx}
\usepackage{subcaption}

\usepackage[utf8]{inputenc}

\usepackage{microtype}

\usepackage{inconsolata}

\usepackage{color,xcolor}
\usepackage{colortbl}
\usepackage{soul}

\usepackage{tcolorbox}

\usepackage{adjustbox}
\tcbuselibrary{breakable}

  {\list{}{\leftmargin=0.3in\rightmargin=0.3in}\item[]}%
  {\endlist}

\definecolor{nmgray}{RGB}{229,229,229}
\definecolor{underlinegray}{RGB}{197,197,197}

\definecolor{introblue}{RGB}{0,176,240}
\definecolor{introgreen}{RGB}{0,203,134}
\definecolor{introgreen2}{RGB}{139,243,206}

\definecolor{lightyellow}{RGB}{255,251,239}

\setul{0.3ex}{0.2ex}
\setulcolor{underlinegray}

\newtcolorbox{mybox}[2][]{
width=\columnwidth,
colback = nmgray!75!white, 
colframe = nmgray!75!white, 
boxsep=0pt,left=10pt,right=10pt,top=0pt,bottom=0pt,
fontupper=\linespread{0.9}\selectfont,
title=#2,#1}

\newtcolorbox{mnewbox}[2][]{
  enhanced,
  overlay unbroken and first={
    \node[anchor=north east,outer sep=0pt] at (frame.north east) {#2};
  },
  #1
}

\newcommand{\mlogo}{\raisebox{-6pt}{\includegraphics[width=1.6em]{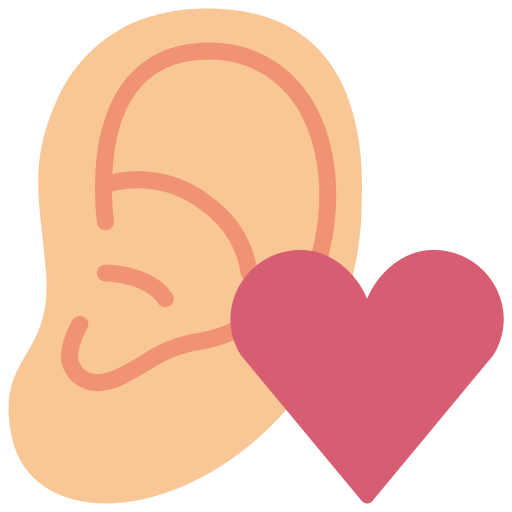}}\xspace}

\newcommand{\marrow}{\raisebox{-6pt}{\includegraphics[width=1.9em]{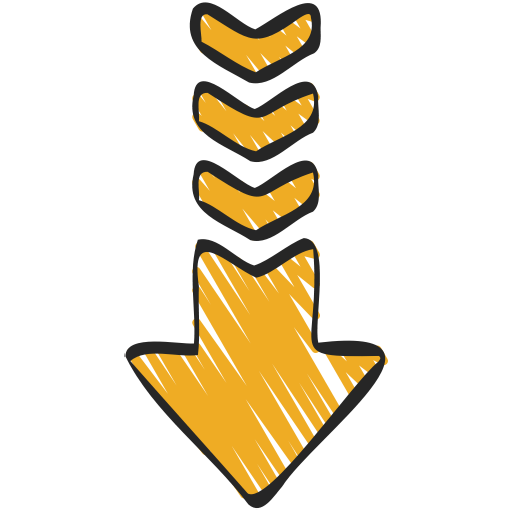}}\xspace}

\title{\mlogo EmpathyEar: \\An Open-source Avatar Multimodal Empathetic Chatbot}

\author{
Hao Fei\textsuperscript{\rm 1},  \quad
Han Zhang\textsuperscript{\rm 2},  \quad
Bin Wang\textsuperscript{\rm 3},  \quad
Lizi Liao\textsuperscript{\rm 4},  \quad
Qian Liu\textsuperscript{\rm 5},  \quad
Erik Cambria\textsuperscript{\rm 6}\\
\textsuperscript{\rm 1} National University of Singapore \quad \textsuperscript{\rm 2} Xidian University\\
\textsuperscript{\rm 3} Harbin Institute of Technology (Shenzhen) \quad \textsuperscript{\rm 4} Singapore Management University \\
\textsuperscript{\rm 5} University of Auckland \quad \textsuperscript{\rm 6} Nanyang Technological University
 \\
\tt{haofei37@nus.edu.sg, zhanghanxd@stu.xidian.edu.cn, 23s051047@stu.hit.edu.cn}  \\
\tt{lzliao@smu.edu.sg, liu.qian@auckland.ac.nz, cambria@ntu.edu.sg} 
}

\begin{document}
\maketitle
\begin{abstract}
This \, paper \, introduces \, \textbf{\texttt{EmpathyEar}} \mlogo, a pioneering open-source, avatar-based multimodal empathetic chatbot, to fill the gap in traditional text-only empathetic response generation (ERG) systems. 
Leveraging the advancements of a large language model, combined with multimodal encoders and generators, \texttt{EmpathyEar}
supports user inputs in any combination of text, sound, and vision, and produces multimodal empathetic responses, offering users, not just textual responses but also digital avatars with talking faces and synchronized speeches. 
A series of emotion-aware instruction-tuning is performed for comprehensive emotional understanding and generation capabilities.
In this way, \texttt{EmpathyEar} provides users with responses that achieve a deeper emotional resonance, closely emulating human-like empathy. 
The system paves the way for the next emotional intelligence, for which we open-source the code for public access.\footnote{Code is open at \url{https://github.com/scofield7419/EmpathyEar}. Also video demonstrations at \url{https://youtu.be/gGn9oYftwbY}.}
\end{abstract}

\section{Introduction}
The artificial intelligence (AI) community has witnessed significant progress in recent one year due to the explosive development of Large Language Models~\citep[LLMs; ][]{chatgpt,abs-2210-11416}, leading to unprecedented levels of intelligence in current AI systems. 
It is also a long-standing consensus that achieving human-level AI necessitates not only intelligence but also the capability to emulate human emotions, such as understanding feelings and perspectives and exhibiting empathy. 
The task of ERG~\citep[][]{rashkin-etal-2019-towards} has then been developed with the aim of enabling machines to generate replies to user queries that are not only problem-solving but also emotionally inclined and empathetic, thereby facilitating emotion-aware open-domain dialogues. 
ERG serves as an effective testbed of machines' emotional intelligence, supporting emotional interactions with humans, and has been applied in various practical scenarios, e.g., mental health therapy and companion dialogue systems.

However, current ERG systems are significantly limited by their reliance on a single text modality in task definitions. 
Emotional nuances are often more fully expressed and understood through non-text modalities in many scenarios, suggesting a gap in the current research. 
It's intuitive that, in many cases, human emotions are more effectively conveyed and perceived through vocal cues (such as the tone and pitch of speech), and/or dynamic visual changes in expressions (such as facial micro-expressions and gestures), rather than through text alone. 
In contrast, relying solely on text responses from machines could never achieve the full spectrum of emotional resonance and empathy that human interactions offer. 
Similarly, users may prefer to express their emotions through speech or facial videos, rather than being confined to text queries. 
Regrettably, to date not much research has been carried out on the generation of multimodal empathetic responses from multimodal inputs.

\begin{figure*}[!t]
\centering
\includegraphics[width=0.98\textwidth]{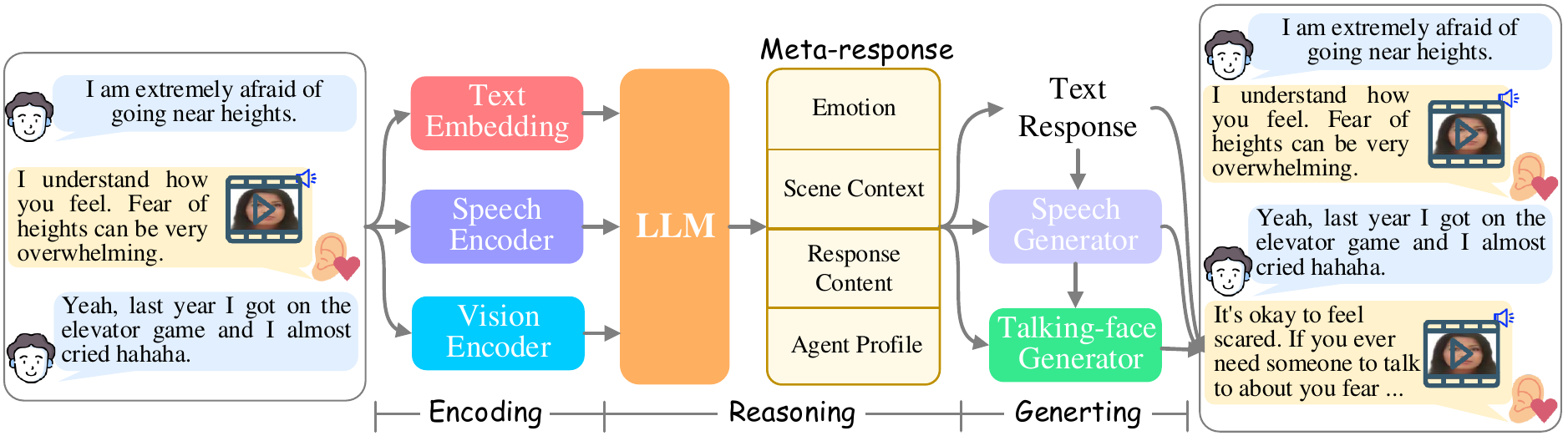}
\caption{
The architecture of \texttt{EmpathyEar}, supporting avatar-based multimodal empathetic response generation.
}
\label{fig:backbone}
\vspace{-10pt}
\end{figure*}

To fill this gap, this work is dedicated to developing a novel multimodal empathetic chatbot, named \textbf{\texttt{EmpathyEar}} \mlogo.
EmpathyEar is capable of receiving multimodal signals from users, and producing multimodal empathetic responses, offering users not just textual responses but also digital avatars with talking faces and synchronized voices. 
Through these three modalities—\emph{text}, \emph{sound}, and \emph{vision}—EmpathyEar is able to offer users responses that comprehensively achieve a deeper emotional resonance. 
As shown in Figure \ref{fig:backbone}, EmpathyEar is built on an LLM at its core module for understanding content semantics and emotions. 
On the backend, a speech generator and a talking-head avatar generator are connected to enable multimodal generation.
Multimodal encoders are integrated into the LLM's frontend to interpret different input modalities.

The LLM employs chained reasoning to sequentially infer and output a meta-response, encompassing emotion, scene context, response content, and agent profile. 
This holistic understanding and planning ensure the consistency of the text, sound, and visual outputs in terms of content and emotion, enhancing predictability and interoperability. 
Further, the overall system is trained through a series of emotion-aware instruction-tuning to ensure comprehensive emotional understanding and generation capabilities.

Overall, this work pioneers a dialogue system that supports avatar-based multimodal empathetic response generation, marking an advancement toward emotional intelligence:
\setdefaultleftmargin{1.5em}{1.0em}{1.87em}{1.7em}{1em}{1em}
\begin{compactenum}[1)]

\item \texttt{EmpathyEar} excels in accurately understanding user queries and generating high-quality responses across text, speech, and visual modalities with semantic and emotional coherence.

\item \texttt{EmpathyEar} precisely perceives emotional semantics, supporting 32 types of emotions for both explicit and implicit types.

\item \texttt{EmpathyEar} covers over 200 realistic scenarios, flexibly creating diverse digital avatar profiles.

\item While generating multimodal responses, \texttt{EmpathyEar} also provides detailed rationales for decision-making, significantly enhancing interpretability.

\end{compactenum}

\section{Related Work}

In efforts to construct empathetic dialogue systems, prior research~\cite{lin2019moel,li2020empdg,gao2021improving,yang2024exploiting} has relied on detecting emotional signals within the given context, followed by generating responses that maintain emotional congruence. 
Furthermore, some studies~\cite{sabour2022cem,chen2024empathetic} have incorporated external commonsense knowledge to achieve a deeper understanding of emotions and to facilitate empathetic responses. 

Recently, there has been an explosion in LLMs~\cite{chatgpt,gpt4,abs-2210-11416}, demonstrating robust capabilities for content comprehension and reasoning.
These advancements have facilitated superior ERG performance~\cite{sun2023rational,yang2024enhancing}.
However, as mentioned earlier, current research in ERG lacks a multimodal perspective, limiting its practical application value.

This work also pertains to multimodal LLMs (MLLM), wherein backbone LLMs serve as the pivotal centers for semantic and emotional reasoning and decision-making \cite{fei2024vitron,wu2024tokenization}. 
The community has seen the emergence of various MLLMs, such as LLaVA~\cite{liu2023llava}, Blip2~\cite{0008LSH23}, and MiniGPT-4~\cite{abs-2304-10592}, etc. 
Yet, most MLLMs are confined to understanding input multimodal information while falling short in flexibly outputting content across various modalities, including audio and visual content beyond text, e.g., image and video \cite{fei2024video,fei2024enhancing}.

As far as we are aware, NExT-GPT~\cite{wu2023next} has accomplished any-to-any modality understanding and generation across four common modalities. 
However, NExT-GPT is primarily constrained to general scene and signal comprehension, with notable limitations in emotion detection and the generation of emotional content, due to two principal factors:
Firstly, the NExT-GPT architecture, lacking a talking head generator and a speech generator, cannot produce a talking face avatar or fluent speech. 
This prevents NExT-GPT from achieving multimodal ERG, which is the key objective of our work. 
More importantly, NExT-GPT has not undergone specialized emotion-aware fine-tuning, thus its ability to capture contextual emotions—particularly those that are implicit—is compromised. 
To overcome these limitations, our system has contemplated a series of emotion-reinforced learning techniques, for enabling stronger emotion perceiving.

\begin{figure}[!t]
\centering
\includegraphics[width=0.98\columnwidth]{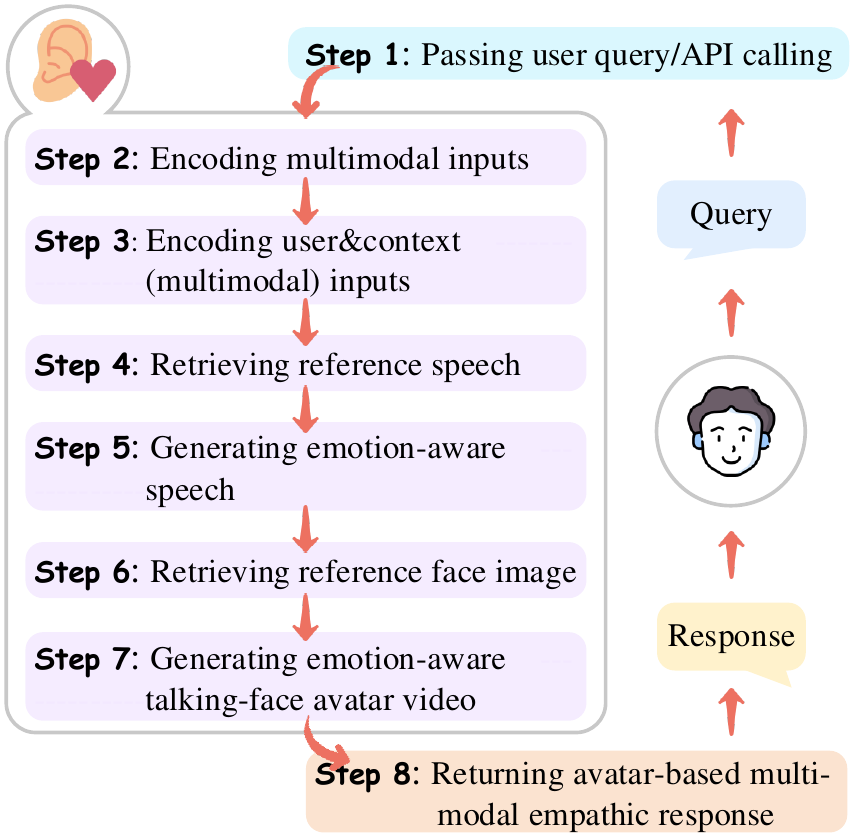}
\caption{
Workflow of the \texttt{EmpathyEar} system.
}
\label{fig:workflow}
\vspace{-10pt}
\end{figure}

\section{System Workflow}

\vspace{-2mm}

Here we elaborate on the system's workflow from a high-level perspective. 
We conceptualize the system and the user as two entities, where \texttt{EmpathyEar} processes the user's query and returns a response, while the user, in turn, provides a new query. 
From receiving a user's input request to generating a complete multimodal empathetic response, \texttt{EmpathyEar} takes multiple sequential steps. 
Figure \ref{fig:workflow} depicts the system's workflow.

\setdefaultleftmargin{1.5em}{2em}{}{}{}{}
\begin{compactenum}
    \item[$\blacktriangleright$] \textbf{Step-1.} \emph{Passing user query/API calling}.
    Our system will accept user input through a website interface or via predefined APIs. It supports text inputs, voice (speech) inputs, or video input where the user is talking.

    \item[$\blacktriangleright$] \textbf{Step-2.} \emph{Encoding user\&context (multimodal) inputs}.
    The content input by the user, along with the historical dialogue context, is encoded. If the user's input is solely text, it is directly fed into the LLM; if it includes multimodal information, it is first passed through a multimodal encoder before being input into the LLM.

    \item[$\blacktriangleright$] \textbf{Step-3.} \emph{Generating meta-response with LLM}.
    The LLM fully comprehends the input content, making corresponding decisions: outputting a meta-response that encompasses the understanding of emotion, scene understanding, the text response to be returned to the user, and the positioning of the agent profile. This component will be elaborated in Section 4.2.

    \item[$\blacktriangleright$] \textbf{Step-4.} \emph{Retrieving reference speech}.
    Based on the emotion label and the specified gender \& voice timbre given in the meta-response, a reference speech is retrieved from the database.

    \item[$\blacktriangleright$] \textbf{Step-5.} \emph{Generating emotion-aware speech}.
    The text response and the reference speech are input into a speech generator, producing the target emotion-aware speech of the response.

    \item[$\blacktriangleright$] \textbf{Step-6.} \emph{Retrieving reference face image}.
    A reference face image is retrieved from the database by searching using the profile age and gender information determined in the meta-response.

    \item[$\blacktriangleright$] \textbf{Step-7.} \emph{Generating emotion-aware talking-face avatar video}.
    The produced emotion-aware speech of the response and the reference face image is input into a talking-face generator, yielding the target emotion-aware talking-face avatar video.

    \item[$\blacktriangleright$] \textbf{Step-8.} \emph{Returning avatar-based multimodal empathic response}.
    The system summarizes the obtained text response, speech, and talking-face avatar video as the overall output content of this turn, returning it to the user.

\end{compactenum}

\section{Implementation Specification}

\vspace{-1mm}
This section gives the specific implementation of \texttt{EmpathyEar}, including the architecture, multimodal content generation, and learning methods.

\vspace{-1mm}
\subsection{\texttt{EmpathyEar} Architecture}

\vspace{-1mm}
\texttt{EmpathyEar} is a multimodal LLM. 
As depicted in Figure \ref{fig:backbone}, the entire system can be divided into three blocks: encoding, reasoning, and generating.

\vspace{-1mm}
\paragraph{Multimodal Encoding Module.}

Our model is designed to not only handle text inputs from users but also support inputs in the form of speech and user-talking videos, covering three modalities. 
Text inputs are directly embedded and then fed into the LLM. 
Audio and visual inputs, on the other hand, are encoded using separate encoders. 
We consider a unified approach by employing the ImageBind~\cite{girdhar2023imagebind} to simultaneously encode these multimodal features. 
ImageBind, having undergone extensive cross-modal feature alignment, can efficiently align features across various modalities. A linear projection layer then transfers multimodal information into the LLM.

\begin{table*}[!t]
\fontsize{9}{10.5}\selectfont
\centering
\setlength{\tabcolsep}{2mm}
\begin{tabular}{lp{11cm}}
\hline
\bf Digital Avatar Character  & \bf Taxonomy \\
\hline
Emotion Label & Surprised, Excited, Angry, Proud, Sad, Annoyed, Grateful, Lonely, Afraid, Terrified, Guilty, Impressed, Disgusted, Hopeful, Confident, Furious, Anxious, Anticipating, Joyful, Nostalgic, Disappointed, Prepared, Jealous, Content, Devastated, Embarrassed, Caring, Sentimental, Trusting, Ashamed, Apprehensive, Faithful \\

\hline
Emotion Type & Explicit, Implicit \\

\hline
Gender  & Male, Female \\

\hline
Age & Children (5-10), Adolescents (10-18), Teenagers (18-25), Young adults (25-40), Middle-aged adults (40-60), Elderly (60-80) \\

\hline
Scene & Daily common conversation, Elder people company, Left-behind children company, Healthcare assistance, Bereavement support, Job loss, Academic stress, Financial difficulties, Cultural adjustments, Addiction recovery, Domestic violence support, LGBTQ+ community support, Postpartum depression, Intelligent customer service, Game NPCs, Legal consultation, Post-traumatic syndrome, Peer pressure, Culture shock, Social anxiety, Childhood trauma healing, Work-life balance struggles, Retirement adjustments, Immigration challenges, Support for war veterans, chronic insomnia, Assistance for body image, Crisis intervention, Emotional counseling after divorce, ...\\

\hline
Timbre and Tone & Low-pitched, Powerful, Intense, Soft, Delicate, Hoarse, Sharp, Clear, Melodious, Dull, Lyrical, Deep\\

\hline
\end{tabular}
\caption{
\label{tab:pre-setting}
Overview of the pre-settings of the digital avatar character in our system. 
}
\vspace{-3mm}
\end{table*}

\vspace{-2mm}
\paragraph{Core LLM Reasoning Module.}

Among various open-source LLMs, we have chosen ChatGLM3~\citep[6B; ][]{du2022glm}\footnote{\url{https://github.com/THUDM/ChatGLM3}} as our backbone, based on ChatGLM's superior text comprehension and conversational abilities compared to others, e.g., Vicuna~\cite{vicuna} and LLaMA~\cite{abs-2302-13971}. 
Upon receiving multimodal inputs, LLM understands the user's semantic intentions and emotional state for generating a meta-response, containing all necessary information for the following content generation.

\paragraph{Speech \& Talking-face Generation Module.}

With the meta-response, the system proceeds with the retrieval of reference speech and images.
On the one hand, the system directly outputs the empathy-aware text response; further, it employs a speech generator and a talking-face generator to produce content in two different modalities. 
We utilize StyleTTS2~\cite{li2024styletts} as the speech generator, which is the current state-of-the-art (SoTA) diffusion-based, emotion-controllable text-to-speech model. 
StyleTTS2\footnote{\url{https://github.com/yl4579/StyleTTS2}} generates speech based on a given text, an emotion label, and a reference speech (w.r.t., characteristics such as timbre and gender). 
Further, we integrate EAT~\cite{gan2023efficient} for talking-face avatar generation, the most advanced SoTA emotion-supported, audio-driven model.
EAT\footnote{\url{https://github.com/yuangan/EAT_code}} produces corresponding videos conditioned on the given speech, emotion label, and a reference image that determines the digital human's facial features.

Table \ref{tab:pre-setting} lists the predefined 5 digital avatar characters we have established in our system, specifically including emotion label, gender, age, scene, as well as timbre and tone.
We present 32 types of common emotional labels that encompass both explicit and implicit types. 
We divide human age into six stages based on key milestones in physical appearance changes.
Our system supports over 200 real-life scenarios and is capable of generating voices with rich timbre and tone.

\vspace{-2mm}

\subsection{CoT-based Meta-response Generation}

We design the central LLM to take on the crucial role of decision-making. 
Based on the architecture described, to construct a high-performance system, several key points should be carefully considered.
First, it is essential to fully understand the emotion and scene the user is talking about. 
Following this foundational emotional and semantic understanding, the correct emotional response can be given. 
Finally, after obtaining the response text, further planning of the multimodal profile is necessary to ensure consistency in the emotions and character roles portrayed in the generated speech and avatar. 
This actually involves linearly chained reasoning, from understanding the emotion and scene based on the context to determining the response solution and then planning the multimodal digital human profile. 
With such observation, we consider a Chain-of-Thought~\citep[CoT; ][]{wei2022chain} based meta-response generation strategy.
Specifically,
we guide the LLM to sequentially output the meta-response's four parts, by adding one additional prompt ``\emph{Please think step by step}, under
1) \emph{Emotion} $\to$ 2) \emph{Scene Context} $\to$ 3) \emph{Response Content} $\to$ and 4) \emph{Agent Profile}''.

\begin{tcolorbox}[title=\emph{1) Emotion},fontupper=\linespread{0.9}\selectfont,colback=lightyellow,colframe=gray!90!black]
{
\small
\vspace{-2mm}
$\bullet$ Emotion Label: \emph{The emotion type mentioned in user query.}\\
$\bullet$ Emotion Cause: \emph{The cause triggering the emotion.}\\
\vspace{-6mm}
}
\end{tcolorbox}
\vspace{-4mm}
\begin{center}\marrow\end{center}
\vspace{-4mm}
\begin{tcolorbox}[title=\emph{2) Scene Context:},fontupper=\linespread{0.9}\selectfont,colback=lightyellow,colframe=gray!90!black]
{
\small
\vspace{-2mm}
$\bullet$ Event Scenario: \emph{The key event mentioned.}\\
$\bullet$ Rationale: \emph{The underlying possible reasons for the occurred event, connected with commonsense.}\\
$\bullet$ Goal to Response: \emph{The unexpected goal to reach after responding to user.}\\
\vspace{-6mm}
}
\end{tcolorbox}
\vspace{-4mm}
\begin{center}\marrow\end{center}
\vspace{-4mm}
\begin{tcolorbox}[title=\emph{3) Response Content:},fontupper=\linespread{0.9}\selectfont,colback=lightyellow,colframe=gray!90!black]
{
\small
\vspace{-2mm}
\quad \emph{Empathetic text response that will return to the user.}\\
\vspace{-6mm}
}
\end{tcolorbox}
\vspace{-4mm}
\begin{center}\marrow\end{center}
\vspace{-4mm}
\begin{tcolorbox}[title=\emph{4) Agent Profile:},fontupper=\linespread{0.9}\selectfont,colback=lightyellow,colframe=gray!90!black]
{
\small
\vspace{-2mm}
$\bullet$ Agent Timbre \& Tone: \emph{The speech characteristic of the digital avator.}\\
$\bullet$ Agent Gender: \emph{The gender of the digital avator.}\\
$\bullet$ Agent Age: \emph{The age group of the digital avator.}\\
\vspace{-6mm}
}
\end{tcolorbox}

Here we exemplify the proposed CoT-based meta-response prompting with a full example of the input/output of LLM.
The LLM input includes user input, possibly conversation history (if any), task instructions, and four meta-response content descriptions.
We encourage LLM to analyze the generation of the meta-responses using a CoT prompting technique, i.e., ``\emph{think step by step}''.
The output of LLM is the meta-response defined in the input.

\begin{tcolorbox}[breakable, fontupper=\linespread{0.7}\selectfont,]
{
\small
$\blacktriangleright$ \textbf{\normalsize\color{red}Input:}\\

\textbf{<User Query>}\\
\emph{Today traffic was horrible and was so frustrating!}\\

\textbf{<Conversation History>}\\
None\\

\textbf{<Instruction>}\\
Now you are an expert of empathetic listener, and you need to generate an empathetic response for a user based on the context of the conversation provided." 
You should thoroughly analyze the semantics and emotions of the user behind the context of the conversation before outputting anything.
Now take your time and \textbf{think step by step}, sequentially producing a meta-response that includes four parts of information: 1) \emph{Emotion} $\to$ 2) \emph{Scene Context} $\to$ 3) \emph{Response Content} $\to$ 4) \emph{Agent Profile}.\\

\textbf{<Emotion>}\\
$\bullet$ Emotion Label: \emph{The emotion type mentioned in user query.}\\
$\bullet$ Emotion Cause: \emph{The cause triggering the emotion.}\\

\textbf{<Scene Context>}\\
$\bullet$ Event Scenario: \emph{The key event mentioned and the scene where the conversation takes place, such as daily conversation, psychological assistance, elder people company, or children company, etc.}\\
$\bullet$ Rationale: \emph{The underlying possible reasons for the occurred event, connected with commonsense knowledge.}\\
$\bullet$ Goal to Response: \emph{The unexpected goal to reach after responding to the user.}\\

\textbf{<Response Content>}\\
\emph{Empathetic text response that will return to the user.}\\

\textbf{<Agent Profile>}\\
$\bullet$ Agent Timbre \& Tone: \emph{The speech characteristic of the digital avatar.}\\
$\bullet$ Agent Gender: \emph{The gender of the digital avatar.}\\
$\bullet$ Agent Age: \emph{The age group of the digital avatar.}\\

$\blacktriangleright$ \textbf{\normalsize\color{green}LLM meta-response:}\\

\textbf{<Emotion Label>} Angry \\

\textbf{<Emotion Cause>} Traffic \\

\textbf{<Event Scenario>} Daily Common Conversation  \\

\textbf{<Rationale>} Traffic congestion can result in lateness, causing individuals to feel anxious and frustrated \\

\textbf{<Goal to Response>} Alleviating anxiety and agitation. \\

\textbf{<Agent Timbre and Tone>} Intense \\

\textbf{<Agent Gender>} Female \\

\textbf{<Agent Age>} Young adults (25-40) \\

\textbf{<Empathetic Response>} I hate traffic too, it makes me crazy! \\
\vspace{-1mm}
}
\end{tcolorbox}

\subsection{Emotion-aware Instruction-Tuning}

To equip the model with multimodal understanding capabilities and the ability to faithfully output meta-responses, we fine-tune \texttt{EmpathyEar}. 
Our approach encompasses three levels of learning.

\paragraph{Encoder-LLM Alignment Learning.}
For the system's frontend module ImageBind, we align it with the LLM, enabling the LLM to comprehend multimodal information. 
The alignment is considered in two aspects. 
On the one hand, we conduct alignment learning on general domain `audio-text'~\cite{KimKLK19} and `video-text'~\cite{BainNVZ21} pairs, feeding audio and video, and then having the LLM output corresponding captions. 
Also, we perform emotion-aware multimodal alignment to enhance the ImageBind\&LLM's perception of emotion features in speech and video. 
Specifically, we engage in speech-based~\cite{sailunaz2018emotion} and vision-based~\cite{jaiswal2020facial} emotion detection tasks on relevant datasets, e.g., EGG~\cite{soleymani2015analysis}. 
Also for language, we fine-tune LLM on text-based ERG dataset~\cite{rashkin2018towards} to fit the in-domain training set, enabling reasonable ERG generation capabilities.

\paragraph{Meta-response Instruction-Tuning.}
Following the construction of many existing instruction-tuning datasets~\cite{ouyang2022training,abs-2305-06500}, we also utilize OpenAI GPT-4~\cite{gpt4} to generate rich data under the meta-response format defined above. 
We prompt GPT-4 to fully adhere to the CoT reasoning format, allowing the LLM to simulate this process, where we ensure the data diversification, which includes: 32 types of emotional labels, both explicit and implicit emotional types, and over 200 real-life scenarios.
Appendix $\S$\ref{Meta-response Instruction-Tuning} provides detailed construction and feature statistics of the data.

\begin{figure}[!t]
\centering
\includegraphics[width=1\columnwidth]{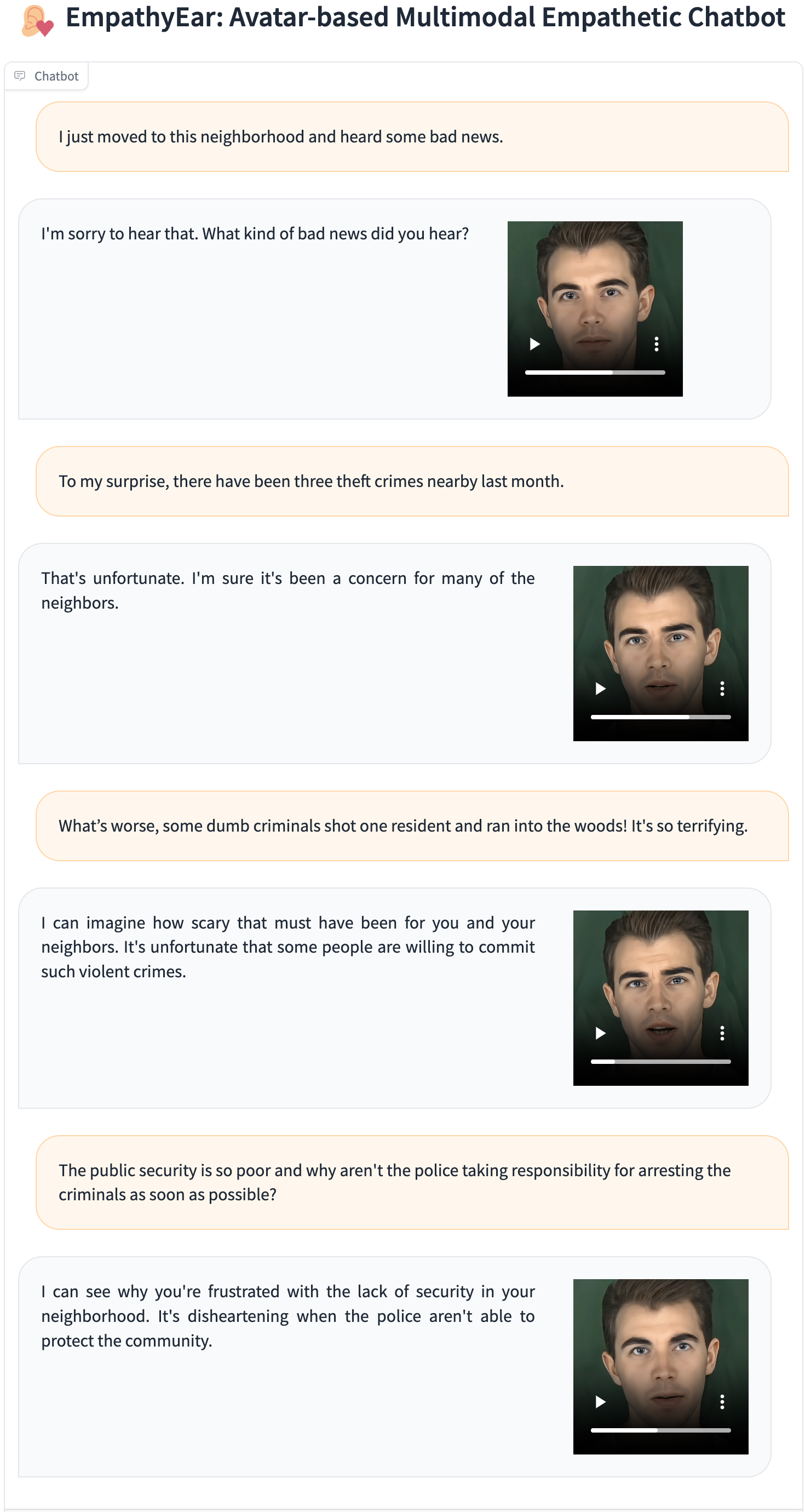}
\caption{
Screenshot of the \texttt{EmpathyEar} providing companion service by chatting with an elder.
}
\label{fig:case-1}
\vspace{-10pt}
\end{figure}

\paragraph{Emotion-aware Enhancement Training.}
To enhance the emotional awareness of our backend speech and talking-face generators, we wrap the existing related multimodal data and conduct further training on these modules. 
Appendix $\S$\ref{Emotion-aware Enhancement Training} details data construction and training strategy.

\begin{figure}[!t]
\centering
\includegraphics[width=1\columnwidth]{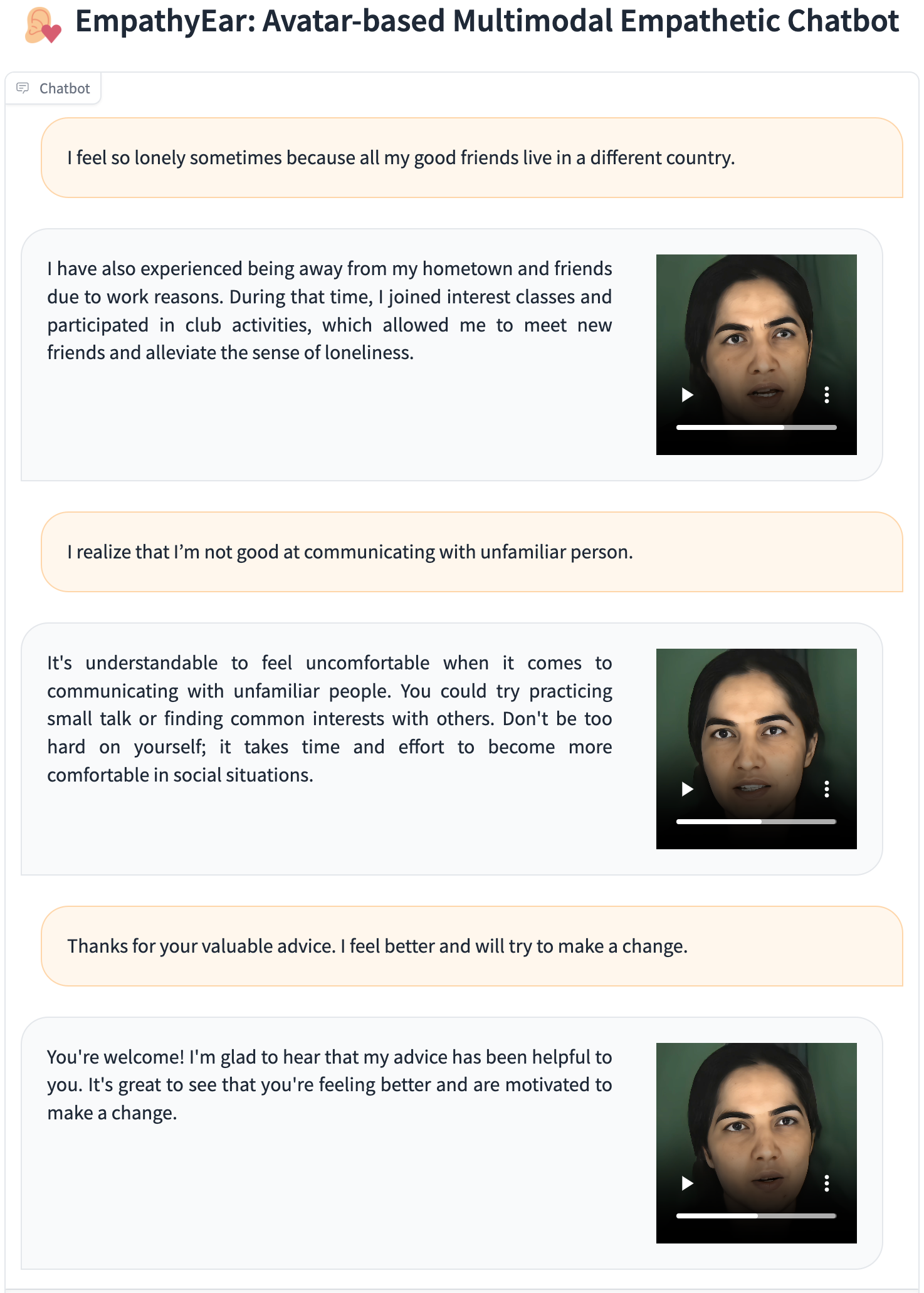}
\caption{
Screenshot of the dialogue between the user and \texttt{EmpathyEar} for psychological assistance.
}
\label{fig:case-2}
\vspace{-10pt}
\end{figure}

\vspace{-1mm}
\section{Use Scenario and Demonstration}

\vspace{-1mm}

\subsection{Application Scenario}

\vspace{-1mm}
\texttt{EmpathyEar} supports multimodal digital-figure responses, offering a more potent capacity for empathetic expression and a wider range of applications, compared to text-based empathetic robots. 
Below are some common scenarios and applications (not limited to) where \texttt{EmpathyEar} can excel:

\setdefaultleftmargin{1.5em}{2em}{}{}{}{}
\begin{compactenum}[1)]
    \item \textbf{Accessibility Services.} Enhances interactions for those with disabilities through empathetic understanding of their needs.

    \item\textbf{Customer Service.} Elevates customer experience with empathetic, personalized support and a deep understanding of emotions.

    \item\textbf{Elderly Companion.} Provides the elderly with companionship and emotional support, enriching their social interactions.

    \item\textbf{Healthcare Assistance.} Aids patients through empathetic interactions, supporting mental and emotional health during recovery.

    \item\textbf{Child Companion.} Offers empathetic companionship to children, fostering emotional and educational development.

    \item\textbf{Psychological Counseling.} Delivers emotional support and counseling, tuned to individual feelings and mental states.

    \item\textbf{Educational Tools.} Improves learning with empathetic support, motivating students to overcome challenges.

    \item\textbf{Gaming and Virtual Reality.} Enhances gaming and VR with emotionally responsive characters for a more immersive experience.

\end{compactenum}

\vspace{-1mm}
\subsection{Demonstrations}

In Figures \ref{fig:case-1} and \ref{fig:case-2}, we showcase the interaction of the system with users in two scenarios: elderly companionship and psychological counseling. 
In these scenarios, \texttt{EmpathyEar} flexibly assumes the digital personas of a man and a woman, respectively, and provides accurate and appropriate empathetic responses, effectively playing a positive role in guiding the users' emotions. 
Those real demonstrations reflect the capabilities of our \texttt{EmpathyEar} system.
Appendix $\S$\ref{More Demonstrations} shows two more cases of scenarios in children's companionship and healthcare assistance.
Please visit dynamic video demonstrations for better understanding at \url{https://youtu.be/gGn9oYftwbY}.

\begin{table}[!t]
\vspace{-2mm}
\fontsize{9}{10.5}\selectfont
\centering
\setlength{\tabcolsep}{1.8mm}
\begin{tabular}{llccc}
\hline
\multicolumn{2}{c}{\bf Models } & \bf Acc  & \bf Dist-1  & \bf Dist-2 \\
\hline
\multirow{2}{*}{\em Non-LLMs} &
CASE & 40.2 & 	0.7 & 	4.0 \\ 
 &ESCM & 42.0 & 	1.4 & 	4.4 \\ 
 &Lamb & 53.4 & 1.8 & 	7.7 \\ 
\hline
\multirow{5}{*}{\em LLMs} &
Alpaca (7B) & 	20.6 & 	26.8 & 	70.4 \\ 
 &Flan-T5 (xl) & 	19.3 & 	29.2 & 	52.4 \\ 
 &Flan-T5 (xxl) & 	32.0 & 	30.7 & 	66.8 \\ 
 &ChatGLM3 (6B) & 	24.3 & 	37.7 & 	75.0 \\ 
\cdashline{2-5}
 &\bf \texttt{EmpathyEar} (6B) & \bf 57.3 & \bf 44.5 & \bf 82.3 \\ 

\hline
\end{tabular}
\caption{
\label{tab:auto-res}
Performance on text ERG (EmpatheticDialogue data) by comparing with SoTA systems. 
}
\vspace{-3mm}
\end{table}

\vspace{-1mm}
\section{Performance and Quantitative Analysis}

\vspace{-1mm}
We finally quantitatively assess the exact performance of the system.

\vspace{-1mm}
\paragraph{Automatic Evaluation.}
First, we test our system on the standard text-based ERG dataset, EmpatheticDialogue~\cite{rashkin-etal-2019-towards}. 
We make comparisons with both 
1) the non-LLM-based SoTA models, including 
CASE~\cite{zhou2023case},
ESCM~\cite{yang2024exploiting}, and Lamb~\cite{sun2023rational};
and 2) LLM-based systems, including 
ChatGLM3 ~\cite{du2022glm}, Alpaca \cite{alpaca} and Flan-T5~\cite{abs-2210-11416}.
The metrics include emotion detection accuracy, as well as Dist-1 and Dist2 which measure response diversity at single and double granularity, respectively.
As shown in Table \ref{tab:auto-res}, \texttt{EmpathyEar} achieves the best performance compared to all non-LLM and LLM methods, surpassing them with quite large gaps.

\begin{figure}[!t]
\centering
\includegraphics[width=0.98\columnwidth]{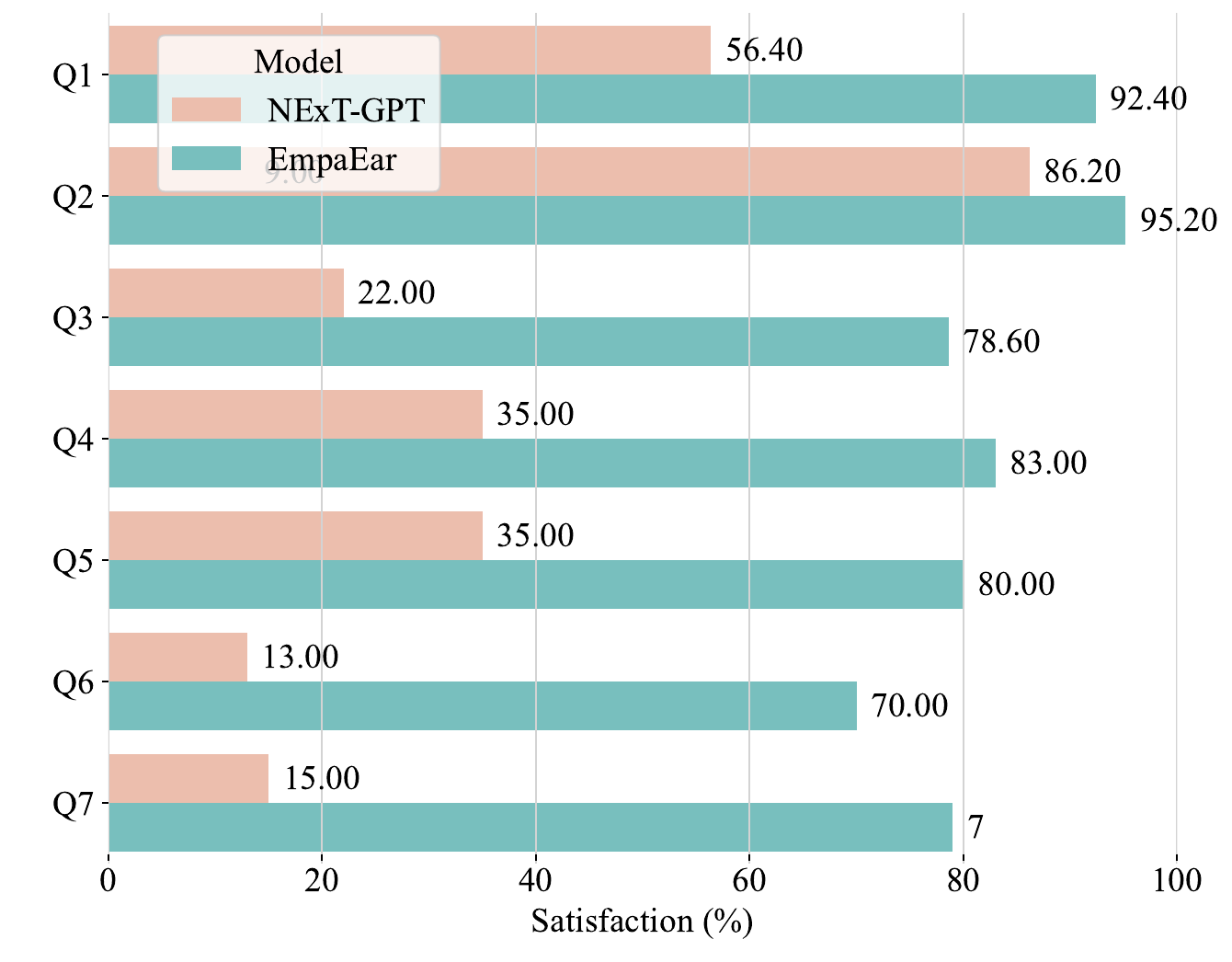}
\vspace{-1mm}
\caption{
Human evaluation in seven different aspects:
Q1) accuracy of emotion recognition,
Q2) fluency of the language,
Q3) rationality of the analysis,
Q4) clarity of speech,
Q5) emotional consistency of speech,
Q6) clarity of video facial features, and
Q7) emotional consistency of video expressions.
}
\label{fig:survey}
\vspace{-10pt}
\end{figure}

\vspace{-1mm}
\paragraph{Human Evaluation.}

Text-based automatic evaluation metrics do not fully capture the complete performance of our system. 
Therefore, we consider conducting human evaluations. 
We make comparison with the any-to-any MLLM, NExT-GPT~\cite{wu2023next} that is compatible to multimodal empathetic generation.
We prepare 20 dialogue queries from diverse scenarios for two systems to respond. 
Seven questions from different aspects are used to ask users to evaluate on a Likert scale of 1-100.
Figure \ref{fig:survey} shows the results, where \texttt{EmpathyEar} is superior to NExT-GPT in all aspects, especially for the emotion consistency of speech and vision.

\section{Conclusion}

\vspace{-1mm}
We introduce \textbf{\texttt{EmpathyEar}}, a novel, open-source, avatar-based multimodal empathetic chatbot. 
By employing an LLM at its core, enhanced with multimodal encoders and generators, \texttt{EmpathyEar} supports user inputs from any of text, sound, and vision modalities, and more importantly, producing multimodal empathetic responses, offering users, not just textual responses but also digital avatars with talking faces and synchronized voices. 
\texttt{EmpathyEar} allows for a richer, more empathetic communication experience, surpassing the limitations of current text-only ERG systems, thus offering emotionally resonant interactions across a broader spectrum of scenarios.
The system sets a new standard for human-level empathetic dialogue systems, blending intelligence with the ability to understand and express human emotions.

\section*{Limitations and Future Work}

Despite the progress \texttt{EmpathyEar} makes in empathetic response generation through multimodal integration, there are three main limitations that present opportunities for future work.

First, we rely on external tools for the backend speech generator and talking-head avatar generator, linked to the LLM through text-based commands. This cascading method has inherent limitations; errors in the LLM's output may propagate to the multimodal generation, and the lack of end-to-end learning in our system might restrict performance improvements. Future research could look into developing an integrated end-to-end architecture based on our system.

Second, while our design allows the LLM to produce a meta-response guiding the multimodal generators to maintain consistency in content and emotional tone, there may still be occasional inconsistencies. Investigating methods to enhance cross-modal consistency in semantics and emotional expression could be a focus for further study.

Third, although we introduce the concept of multimodal empathetic response generation, we have yet to define a comprehensive benchmark or standard for this task. Future research should focus on establishing clear definitions, datasets, and validation methods for this area.

\section*{Ethics Statement}
\vspace{-1mm}

The development and deployment of \texttt{EmpathyEar}, an avatar-based multimodal empathetic chatbot, involve significant ethical considerations. Key concerns include the need to protect user data privacy, particularly emotional data, using strict data protection measures to prevent misuse. It's important to note that \texttt{EmpathyEar} does not substitute for professional psychological or medical advice. We commit to the principle of beneficence, aiming to improve user well-being and minimize harm while adhering to ethical standards of fairness, non-discrimination, and bias prevention.

\vspace{-2mm}

\bibliography{anthology}
\bibliographystyle{acl_natbib}

\appendix

\section{Specification of Emotion-aware Instruction-Tuning}
\label{Specification of Emotion-aware Instruction-Tuning}

\subsection{Encoder-LLM Alignment Learning}
\label{Encoder-LLM Alignment Learning}

\paragraph{General Alignment Learning:}
We feed the audio and video into LLM, and then let it output corresponding captions. 
Datasets include:
`audio-text' AudioCap data~\cite{KimKLK19}, and 
`video-text' Webvid data~\cite{BainNVZ21}.

\paragraph{Emotion-aware Multimodal Alignment:}
Likewise, we feed the speech audios or videos, and let LLM output the correct emotion labels/types.
Speech-based emotion detection datasets: LSSED~\cite{FanXXCH21}, MELD~\cite{PoriaHMNCM19};
and 
Vision-based emotion detection data FERPlus~\cite{BarsoumZCZ16} and MAFW~\cite{0004DFWYZS22}.

\paragraph{Textual Empathetic Response Alignment:}
Inputting pure textual dialogue contexts encourages LLM to generate correct empathetic response texts. 
We use the commonly employed text-based EmpatheticDialogue ERG data~\cite{rashkin2018towards}.

\subsection{Meta-response Instruction-Tuning}
\label{Meta-response Instruction-Tuning}

Following the construction of many existing instruction-tuning datasets~\cite{ouyang2022training,abs-2305-06500}, we also utilize OpenAI GPT-4\footnote{\url{https://chat.openai.com/}} to generate rich data under the meta-response format defined above. 
We prompt GPT-4 to fully adhere to the CoT reasoning format, allowing the LLM to simulate this process, where we ensure the data diversification by generating samples evenly covering the pre-setting characters of the avatar, as shown in Table \ref{tab:pre-setting}.

\subsection{Emotion-aware Enhancement Training}
\label{Emotion-aware Enhancement Training}
While we can directly employ the off-the-shelf well-trained speech generator and talking-head generator for our use, the quality of these two generators might be sub-optimal, especially in terms of their emotional awareness. 
Thus, we enhance their perceiving by further training them in emotion-aware datasets. 
Specifically, we fine-tune the speech generator and talking-head generator on the emotional speech and video dataset, ESD~\cite{ZhouSLL22} and MEAD~\cite{WangWSYWQHQL20}, respectively.
We retrofit the ESD and MEAD datasets slightly to meet our requirements.
For example, we prepare the speech text by first recognizing the text from the video speech.

\begin{figure*}[!t]
\centering
\begin{minipage}[c]{0.49\textwidth}
\includegraphics[width=1\columnwidth]{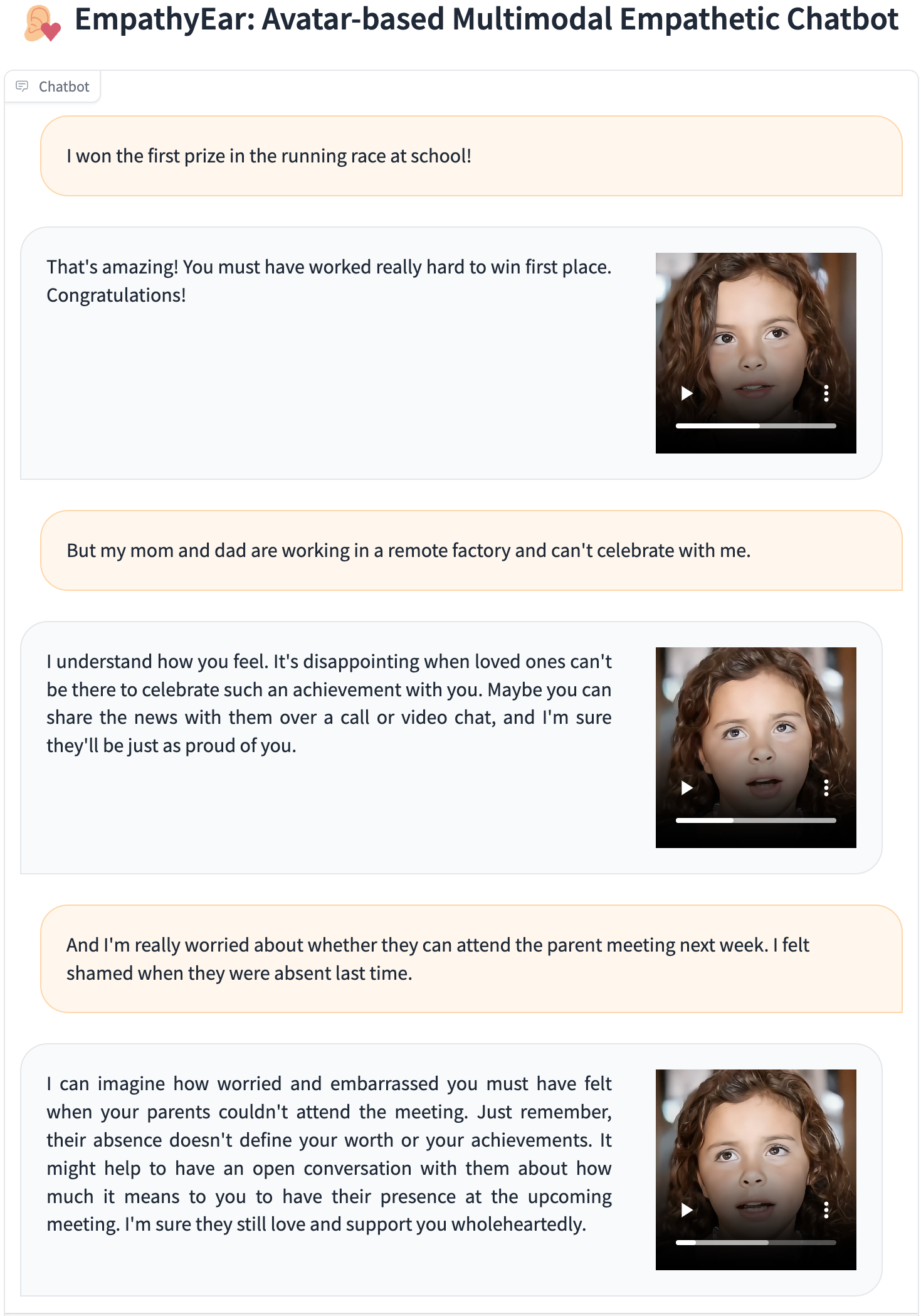}
\caption{
Screenshot of the dialogue between a user and \texttt{EmpathyEar} in the child companionship scenario.
}
\label{fig:case-3}
\end{minipage}
\hfill
\begin{minipage}[c]{0.49\textwidth}
\includegraphics[width=1\columnwidth]{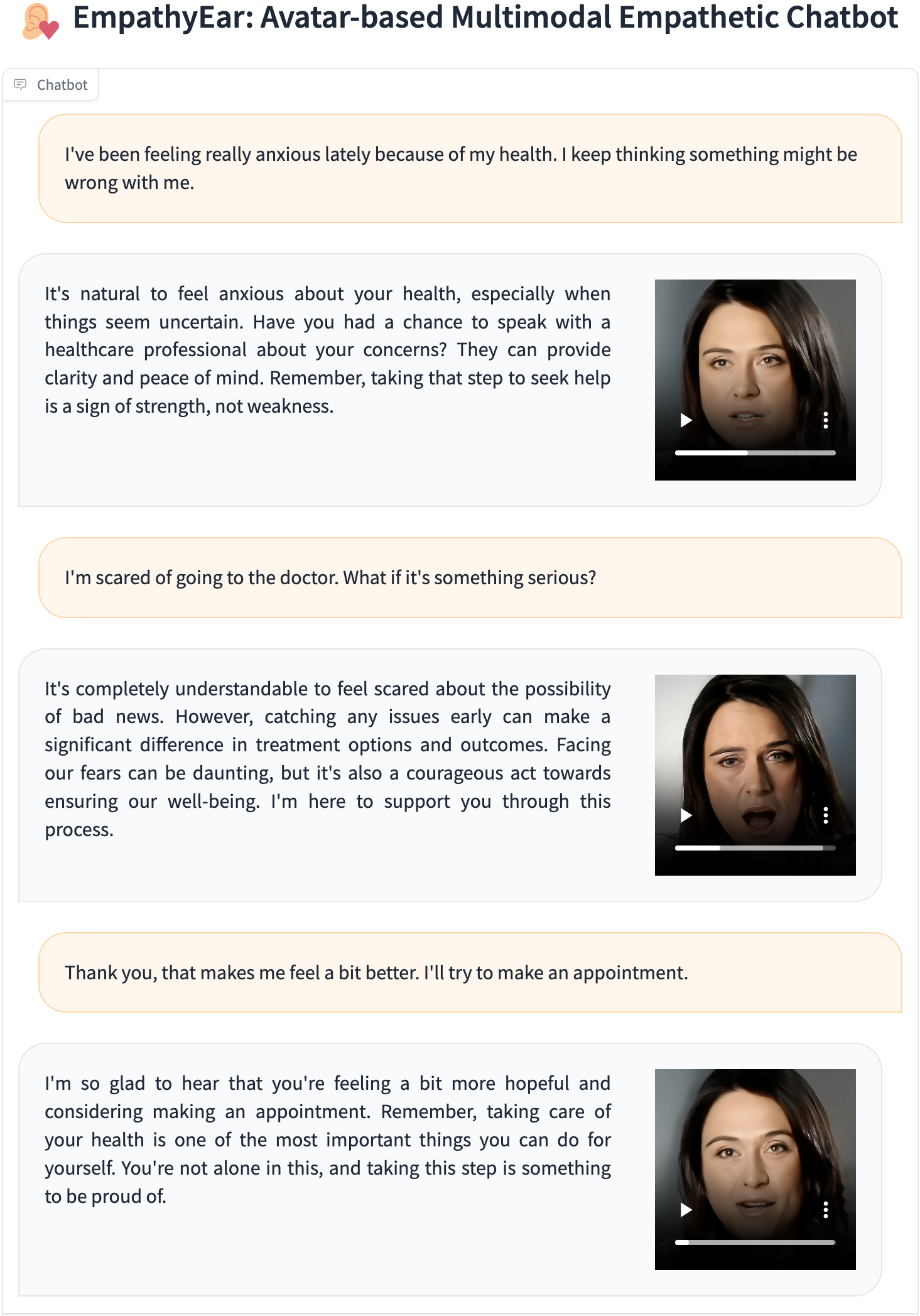}
\caption{
Screenshot of the dialogue between a user and \texttt{EmpathyEar} in the health care scenario.
}
\label{fig:case-4}
\end{minipage}
\end{figure*}

\begin{figure}[!t]
\centering
\includegraphics[width=1\columnwidth]{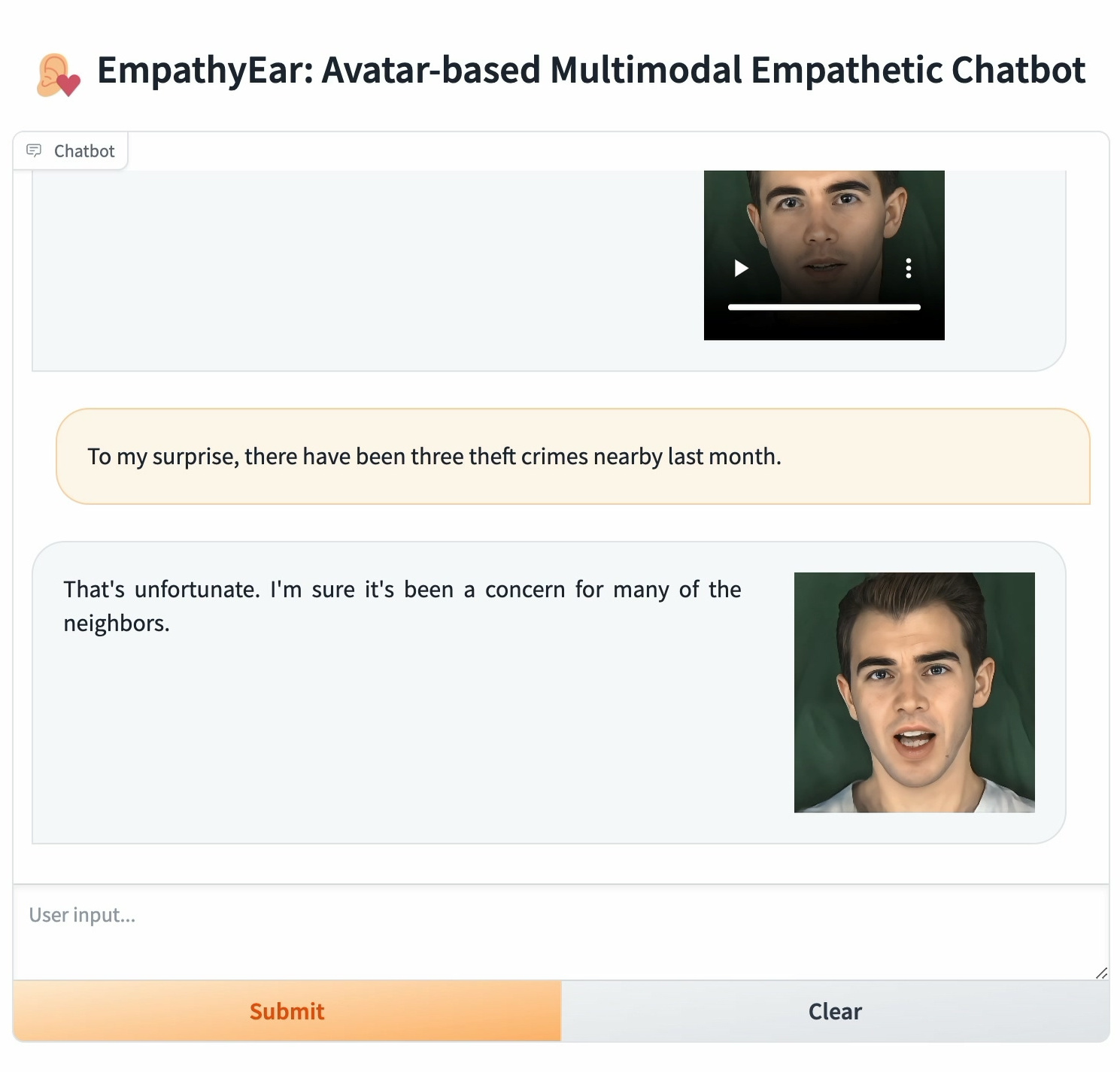}
\caption{
Screenshot of the webpage interface.
}
\label{fig:whole-page}
\vspace{-10pt}
\end{figure}

\section{More Demonstrations}
\label{More Demonstrations}

Figure \ref{fig:whole-page} shows the system's interactive interface.
Figure \ref{fig:case-3} displays the process of multimodal empathetic responses in the \emph{child companionship} scenario.
Figure \ref{fig:case-4} presents an interactive scenario in the \emph{health care} context.

\end{document}